\documentclass[conference]{IEEEtran}
\IEEEoverridecommandlockouts
\usepackage{graphicx}
\graphicspath{{./myfigures/}}
\usepackage{amsmath}
\usepackage{amssymb}
\usepackage{bm}
\usepackage{subfigure}
\usepackage{algorithm}  
\usepackage{algorithmicx}  
\usepackage{algpseudocode}
\usepackage{cite}
\usepackage{color}
\usepackage{siunitx}
\newcommand{\ie}{\textit{i}.\textit{e}.}

\def\dif{\mathop{}\hphantom{\mskip-\thinmuskip}\mathrm{d}}%
\let\daccent\d
\let\d\relax
\newcommand\d{\ifmmode\dif\else\expandafter\daccent\fi}

\begin{document}

\title{Joint Hybrid Beamforming and Artificial Noise Design for Secure Multi-UAV ISAC Networks
\thanks{This work was supported in part by the National Natural Science Foundation of China under Grant 62301600 and 62101560.}
}
\author{\IEEEauthorblockN{1\textsuperscript{st} Runze Dong}
	\IEEEauthorblockA{\textit{School of Information and Navigation} \\
		\textit{Air Force Engineering University}\\
		Xi'an, China \\
		drzaxx@buaa.edu.cn}
	\and
	\IEEEauthorblockN{2\textsuperscript{nd} Buhong Wang}
	\IEEEauthorblockA{\textit{School of Information and Navigation} \\
		\textit{Air Force Engineering University}\\
		Xi'an, China \\
		wbhgroup@aliyun.com}
	\and
	\IEEEauthorblockN{3\textsuperscript{rd} Cunqian Feng}
	\IEEEauthorblockA{\textit{School of Air Defense and Antimissile} \\
	\textit{Air Force Engineering University}\\
		Xi'an, China \\
		fengcunqian@sina.com}
		
	\and
	\IEEEauthorblockN{4\textsuperscript{th} Jiang Weng}
	\IEEEauthorblockA{\textit{School of Information and Navigation} \\
	\textit{Air Force Engineering University}\\
		Xi'an, China \\
		wengjiang858@163.com}
	
	\and
	\IEEEauthorblockN{5\textsuperscript{th} Chen Han}
	\IEEEauthorblockA{\textit{Sixty-Third Research Institute} \\
		\textit{National University of Defense Technology}\\
		Nanjing, China \\
		chenhan2017lgd@163.com}

	\and
	\IEEEauthorblockN{6\textsuperscript{th} Jiwei Tian}
	\IEEEauthorblockA{\textit{School of Air Traffic Control and Nav.} \\
		\textit{Air Force Engineering University}\\
		Xi'an, China \\
		tianjiwei2016@163.com}
}
\maketitle

\begin{abstract}
Integrated sensing and communication (ISAC) emerges as a key enabler for next-generation applications such as smart cities and autonomous systems. Its integration with unmanned aerial vehicles (UAVs) unlocks new potentials for reliable communication and precise sensing in dynamic aerial environments. However, existing research predominantly treats UAVs as aerial base stations, overlooking their role as ISAC users, and fails to leverage large-scale antenna arrays at terrestrial base stations to enhance security and spectral efficiency. This paper propose a secure and spectral efficient ISAC framework for multi-UAV networks, and a two-stage optimization approach is developed to jointly design hybrid beamforming (HBF), artificial noise (AN) injection, and UAV trajectories. Aiming at maximizing the sum secrecy rate, the first stage employs Proximal Policy Optimization (PPO) to optimize digital beamformers and trajectories, and the second stage decomposes the digital solution into analog and digital components via low-complexity matrix factorization. Simulation results demonstrate the effectiveness of the proposed framework compared to benchmark schemes.
\end{abstract}


\section{Introduction}
With the growing demand for ubiquitous connectivity, immersive communication, and high spectrum utilization and potential application scenarios such as smart city, industrial automation, and autonomous driving for the next generation communication networks, integrated sensing and communication (ISAC) has brought about widespread attention \cite{RN1139, RN1095}. This integration is particularly critical in dynamic and complex scenarios where efficient resource allocation and real-time adaptability are significant. Unmanned aerial vehicles (UAVs) are renowned for their inherent mobility, rapid deployment capability, and favorable line-of-sight (LoS) propagation conditions. These qualities have led to the emergence of UAVs as flexible and agile platforms capable of enhancing such ISAC-enabled networks, which makes them ideal candidates for applications ranging from coverage expansion and public safety to precision monitoring and disaster response. Therefore, the convergence of ISAC and UAV represents a promising path towards next-generation wireless systems that are both spectral efficient and event aware \cite{RN1094, RN1087, RN1028}.

ISAC enabled UAV systems have garnered significant research attention, with prior work focusing on trajectory and beamforming co-design \cite{RN873} and resource allocation under performance trade-offs \cite{RN1028}. Despite these advances, few works focus on the security performance \cite{RN1029, RN1132}. An intelligent reflecting surface (IRS) was introduced in \cite{RN1029} to assist the secure transmission of UAV, and its transmit power, user scheduling, and trajectory were jointly optimized to maximize the average achievable rate. In \cite{RN1132}, alternating optimization (AO) strategy was employed to design the trajectory and beamforming of an UAV base station.

In previous reviewed works UAVs are predominantly deployed as ISAC base stations, with a primary focus on optimizing network parameters through conventional methods such as convex optimization and AO. However, an equally important and practical application scenario arises when UAVs act as legitimate aerial users served by terrestrial ISAC base stations, such as in real-time surveillance systems, emergency response coordination, where secure and reliable airborne communication is critical. Furthermore, the use of large-scale antenna arrays in such configurations can more effectively exploit spatial degrees of freedom, significantly enhancing physical layer security through precise beamforming and artificial noise (AN) injection.

The increasing complexity and capacity demands of wireless networks have positioned DRL as a critical tool for addressing non-convex optimization challenges in dynamic communication networks. An millimeter wave (mmWave) heterogeneous network including diverse multiple access schemes was considered in \cite{RN1144}, and DRL was employed to tackle a multi-objective optimization problem aimed at maximizing user fairness while minimizing transmit power. Similarly, \cite{RN1036} investigated an mmWave vehicular network scenario where DRL was leveraged to enhance the secrecy capacity of all target vehicles under energy consumption constraint. Furthermore, with the emergence of artificial intelligence generated content (AIGC) services, multiple UAVs were integrated as agile service providers in \cite{RN1104}, and a diffusion model-enhanced DRL framework was developed to optimize computation task offloading and resource management policies. These studies collectively demonstrate the adaptability and effectiveness of DRL in handling complex, high-dimensional optimization problems across various wireless network architectures.

Motivated by the aforementioned challenges and opportunities, this paper proposes a joint beamforming and AN design to simultaneously enhance communication security and spectral efficiency in dynamic multi-UAV ISAC networks. As the pricing budget for configuring each antenna with a dedicated radio frequency (RF) chain, the hybrid beamforming (HBF) framework is adopted \cite{RN1142}. The objective is to maximize the sum secrecy rate across all legitimate UAVs, by jointly optimizing the hybrid beamformer, AN, and trajectories of legitimate UAVs. A two-stage optimization framework is developed to balance performance and computational tractability. In the first stage, the digital beamformer and UAV trajectories are optimized using Proximal Policy Optimization (PPO), chosen for its favorable convergence properties and training stability. The second stage employs an AO-based method with low-complexity to decompose the fully-digital solution into its analog and digital components under power and constant modulus constraints.

The remaining part of this paper is organized as follows. Section II presents the system and signal model of the multi-UAV ISAC network, the considered sum secrecy rate maximization problem is formulated and analyzed in Section III. After that, Section IV investigate the two-stage optimization framework with DRL and matrix factorization incorporated. Simulation experiments are demonstrated in Section V, Section VI concludes this paper.

\textit{Notations:} Throughout this paper, matrices and vectors are denoted by bold-faced upper case and lower case letters, while scalar quantities are denoted by italic letters. $\bm x^T$ and $\bm x^H$ denote the transpose and conjugate transpose, $\left[x\right]^+$ demonstrate to take the maximum value between $x$ and 0, $\bm X_{i,j}$ means the element in the $i$-th row and $j$-th column of matrix $\bm X$. $\bm X^{\dagger}$ is the Moore-Penrose pseudo-inverse of matrix $\bm X$.
\section{System and Signal Model}
\begin{figure}[t]
	\centering
	\includegraphics[width=2.5in]{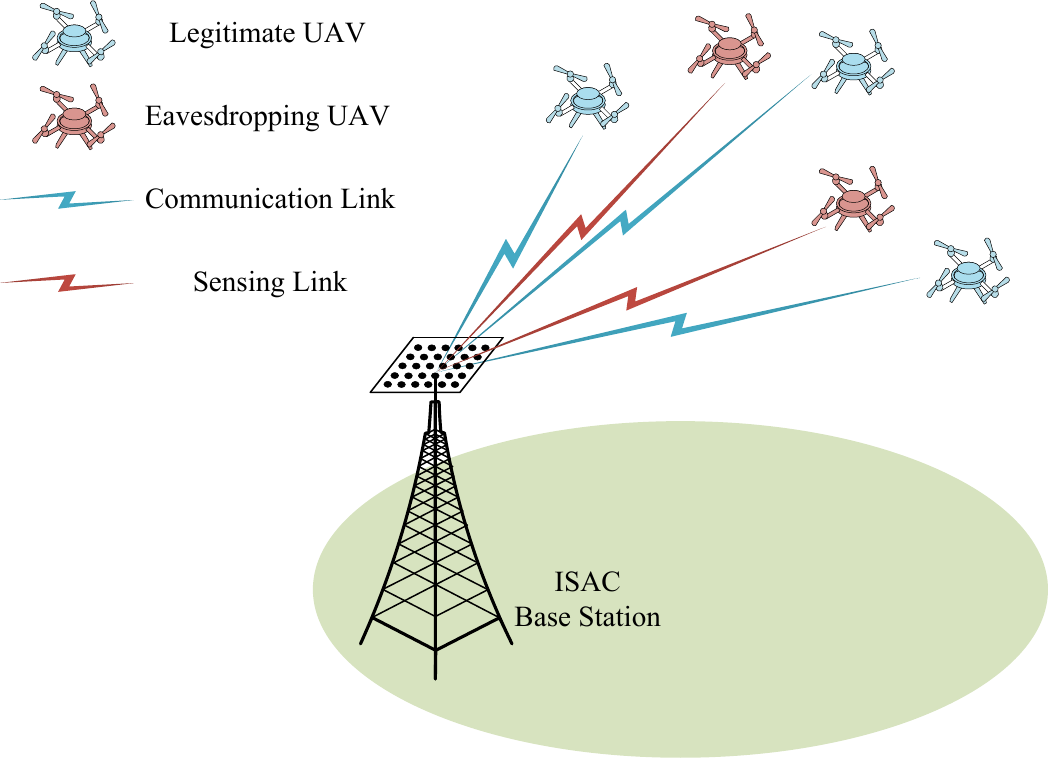}
	\caption{The Multi-UAV uplink transmission network with ISAC.}
	\vspace{-0.3em}
	\label{SysFig}
\end{figure}
\subsection{System Model}
The considered system model, depicted in Fig. \ref{SysFig}, features a multi-UAV uplink ISAC network. A terrestrial ISAC base station employs space division multiple access (SDMA) to transmit confidential information to $L$ legitimate UAVs simultaneously. This approach enhances transmission efficiency and reduces task planning duration. However, $E$ eavesdropper UAVs located near the legitimate UAVs attempt to intercept the communication. This scenario abstractly models a practical use case where a base station delivers mission-critical data to a UAV swarm prior to takeoff.

The ISAC base station is equipped with a uniform planar array (UPA) comprising $N_t$ antennas and $N_{RF}$ RF chains, where $L\leqslant N_{RF}\leqslant N_t$. All UAVs are assumed to be equipped with a single antenna. The total transmit duration $T$ is divided into $N$ sufficient short time slots such that $T=N\Delta_t$, where $\Delta_t$ is the time interval. The position of the $l$-th legitimate UAV in the $n$-th time slot is denoted by $\bm c_l[n]=\left[x_l[n], y_l[n],z_l\right]^T$. It is noteworthy that each legitimate UAV is assumed to maintain a fixed flight altitude to avoid collisions with others. In contrast, all eavesdropper UAVs are assumed to be static. In addition to transmitting confidential information to legitimate UAVs, the ISAC base station proactively broadcasts AN, which serves the dual purpose of jamming potential eavesdropping UAVs and providing a sensing signal to estimate their locations.
\subsection{Communication Model}
Prior to transmission, the generated confidential signal and AN are processed by HBF, which involves firstly applying digital beamforming via RF chains and then analog beamforming through phase shifter. Therefore, the composite transmitted signal from the ISAC base station could be expressed as\vspace{-0.3em}
\begin{equation}
	\bm x[n]=\bm F_{RF}[n]\left(\textstyle\sum_{l=1}^{L}\bm f_{BB,l}[n]s_l[n]+\bm w[n]z[n]\right),\label{eq1}\vspace{-0.3em}
\end{equation}
where $\bm F_{RF}[n]\in\mathbb{C}^{N_{t}\times N_{RF}}$ is the analog beamformer, $\bm f_{BB,l}[n]\in\mathbb{C}^{N_{RF}\times 1}$ is the digital beamformer for signal $s_l[n]$. To facilitate comprehension, define $\bm F_{BB}[n]\triangleq\left[\bm f_{BB,1}[n],\bm f_{BB,2}[n],\ldots,\bm f_{BB,L}[n]\right]$ as the overall digital beamformer for all confidential signals, which will leading (\ref{eq1}) to\vspace{-0.4em}
\begin{equation}
	\bm x[n]=\bm F_{RF}[n]\left(\bm F_{BB}[n]\bm s[n]+\bm w[n]z[n]\right),\label{eq2}\vspace{-0.3em}
\end{equation}
in which $\bm s[n]=\left[s_1[n],s_2[n],\ldots,s_L[n]\right]^T$ denoting the $L$ confidential data streams corresponding to $L$ legitimate UAVs, $z[n]$ is the AN and $\bm w[n]\in\mathbb{C}^{N_{RF}\times 1}$ is the digital beamformer for it. It is assumed that the confidential signal $\bm s[n]$ and AN are statistically independent, with $\mathbb{E}[\bm{s}[n]\bm{s}^H[n]] = \bm{I}_L$ and $\mathbb{E}[|z[n]|^2] = 1$. As such, the received signals at the $l$-th legitimate UAV and the $e$-th eavesdropping UAV could be expressed as\vspace{-0.2em}
\begin{subequations}\label{2}
	\begin{equation}
		y_l[n]=\bm h_l^H[n]\bm x[n]+n_l[n],\label{received_B}
	\end{equation}
	\begin{equation}
		y_e[n]=\bm h_e^H[n]\bm x[n]+n_e[n],\label{received_E}
	\end{equation}
\end{subequations}
where $\bm h_l[n]\in\mathbb C^{N_t\times 1}$ and $\bm h_e[n]\in\mathbb C^{N_t\times 1}$ represent the channel gains from the ISAC base station to the $l$-th legitimate UAV and the $e$-th eavesdropper UAV respectively. The additive white Gaussian noises (AWGN) at the corresponding receivers is denoted by $n_l[n] \sim \mathcal{CN}(0, \sigma_l^2)$, $n_e[n] \sim \mathcal{CN}(0, \sigma_e^2)$. Following the simplified mmWave channel model \cite{RN948}, the channel gains $\bm h_{i},i\in\{l,e\}$ is given by\vspace{-0.2em}
\begin{equation}
	\bm h_i=\beta_0\alpha_0\mathbf{a}(\phi_i,\theta_i),\vspace{-0.2em}
\end{equation}
where $\beta_0=\frac{c}{4\pi f_c}d_i^{-\kappa}$ is the channel gain coefficients, $d_i$ is the distance between the ISAC base station and node $i$, $\alpha_0\sim \mathcal{CN}(0, 1)$ is the small-scale fading gain, $\mathbf{a}(\phi,\theta)$ is the steering vector corresponding to the UPA on the ISAC base station\vspace{-.3em}
\begin{equation}
	\begin{split}		\mathbf{a}(\phi,\theta)=&\frac{1}{\sqrt{N_t}}\left[1,\ldots,e^{-j\pi\sin\theta(n_x\cos\phi+n_y\sin\phi)},\right.\\&\quad\left.\ldots,e^{-j\pi\sin\theta((N_x-1)\cos\phi+(N_y-1)\sin\phi)}\right]^T,\vspace{-0.3em}
	\end{split}
\end{equation}
which is oriented parallel to the $x-o-y$ plane and following the half wavelength spacing assumption. $N_x$ and $N_y$ is the antenna number along the $x$-axis and $y$-axis respectively.
\addtolength{\topmargin}{0.01in}

Substituting the transmitted signal model (\ref{eq2}) into (\ref{received_B}) and (\ref{received_E}), and assuming the $e$-th eavesdropper UAV is interested in the confidential signal corresponding to the $l$-th legitimate UAV, we have\vspace{-0.5em}
\begin{subequations}
	\begin{equation}
		\begin{split}
			y_l[n]\!&=\!\underbrace{\bm h_l^H\bm F_{RF}[n]\bm f_{BB,l}[n] s_l[n]}_{\text{Desired Signal}}\!+\!\underbrace{\sum_{j\neq l} ^L\bm h_l^H\bm F_{RF}[n]\bm f_{BB,j}[n] s_j[n]}_{\text{Multi-User Interference}}\\&+\underbrace{\bm h_l^H\bm F_{RF}[n]\bm w[n]\bm z[n]}_{\text{AN Interference}}+n_l[n],\raisetag{5ex}
		\end{split}
	\end{equation}\vspace{-0.5em}
	\begin{equation}
		\begin{split}
			y_{e,l}[n]\!&=\!\underbrace{\bm h_e^H\bm F_{RF}[n]\bm f_{BB,l}[n] s_l[n]}_{\text{Signal in Interest}}\!+\!\underbrace{\sum_{j\neq l} ^L\bm h_e^H\bm F_{RF}[n]\bm f_{BB,j}[n] s_j[n]}_{\text{Interference}}\\&+\underbrace{\bm h_e^H\bm F_{RF}[n]\bm w[n]\bm z[n]}_{\text{Jamming by AN}}+n_e[n],\raisetag{5ex}\vspace{-0.3em}
		\end{split}
	\end{equation}
\end{subequations}
and the received signal-to-interference plus noise ratio (SINR) could be given by\vspace{-0.3em}
\begin{subequations}
	\begin{equation}
		\gamma_l[n]\!=\!\frac{\left|\bm h_l^H\bm F_{RF}[n]\bm f_{BB,l}[n]\right|^2}{\sum_{j\neq l} ^L\left|\bm h_l^H\bm F_{RF}[n]\bm f_{BB,j}[n]\right|^2\!+\!\left|\bm h_l^H\bm F_{RF}[n]\bm w[n]\right|^2\!+\!\sigma_l^2},
	\end{equation}
	\begin{equation}
		\gamma_{e,l}[n]\!=\!\frac{\left|\bm h_e^H\bm F_{RF}[n]\bm f_{BB,l}[n]\right|^2}{\sum_{j\neq l} ^L\left|\bm h_e^H\bm F_{RF}[n]\bm f_{BB,j}[n]\right|^2\!+\!\left|\bm h_e^H\bm F_{RF}[n]\bm w[n]\right|^2\!+\!\sigma_e^2}.
	\end{equation}
\end{subequations}

Following that, the achievable secrecy rate of the $l$-th legitimate UAV in time slot $n$ could be give by
\begin{equation}
	R_{s,l}[n]=[R_l[n]-\underset{e\in E}{\text{max}}R_{e,l}[n]]^+,
\end{equation}
where $R_l[n]=\log_2\left(1+\gamma_l[n]\right)$ is the achievable rate of the legitimate UAV $l$, $R_{e,l}[n]=\log_2\left(1+\gamma_{e,l}[n]\right)$ is the achievable of the eavesdropper UAV $e$ while wiretap legitimate UAV $l$.

\subsection{Sensing Model}
In this work we consider a holistic sensing paradigm that leverages the entire transmitted waveform. The sensing performance is evaluated based on the complete transmitted signal $\bm x[n]$, which synergistically combines the confidential signals and AN. The covariance matrix of the composite signal as the key metric for sensing performance is given by
\begin{equation}
	\begin{split}
		\bm R_{\bm x}[n]=&\mathbb{E}\{\bm x[n]\bm x^H[n]\}\\=&\bm F_{RF}[n]\!\left(\bm F_{BB}[n]\bm F_{BB}^H[n]\!+\!\bm w[n]\bm w^H[n]\right)\!\bm F_{RF}^H[n],\!
	\end{split}		
\end{equation}
and the resulting sensing beampattern at elevation angle $\phi$ and azimuth angle $\theta$ can be expressed as
\begin{equation}
	P_{\phi,\theta}[n]=\mathbf{a}^H(\phi,\theta)\bm R_{\bm x}[n]\mathbf{a}(\phi,\theta).
\end{equation}

The fundamental accuracy of target parameter estimation (e.g., angle, range) is bounded by the signal-to-noise ratio (SNR) of the reflected echo, which is a direct function of the incident power density. Consequently, the spatial distribution of transmitted power, defined by the beampattern, serves as a critical and deterministic proxy for predicting sensing performance. Accordingly, a sensing performance threshold is preset for each eavesdropper UAV, which means the following performance constraint should be ensured
\begin{equation}
	P_{\phi_e,\theta_e}[n]d_e^{-2}\geqslant\Gamma_e,\forall n.\label{SenCon}
\end{equation}

After the communication and sensing models are established, we can turn to the sum secrecy rate maximization problem formulation, which will be demonstrated in the next section.
\section{Problem Formulating and Analysis}
This work aims to maximize the sum secrecy rate across all legitimate UAVs while guaranteeing satisfactory sensing performance towards all eavesdropper UAVs via the joint optimization of HBF and trajectories of legitimate UAVs. Achieving this objective requires a fundamental trade-off: the limited spatial degrees of freedom afforded by HBF must be optimally allocated. Specifically, these resources are used to simultaneously shape the constructive and destructive interference patterns of both the confidential signal and AN throughout the propagation environment, which finally lead to the following optimization problem
\begin{subequations}
	\begin{equation}
		\underset{\bm F_{RF},\bm F_{BB}, \bm w,\bm c_l}{\mathop{\text {max}}}\sum_{n=1}^{N}\sum_{l=1}^{L}R_{s,l}[n] \notag
	\end{equation}
	\vspace{-1.3em}
	\begin{align}
		s.t.\ &\left\|\bm F_{RF}[n]\bm F_{BB}[n]\right\|_F^2+\left\|\bm F_{RF}[n]\bm w[n]\right\|_2^2\leqslant P_t,\forall n, \label{c1}\\
		&\left|\bm F_{RF}[n]_{i,j}\right|=1/\sqrt{N_t},\forall i,j,n, \label{c2}\\
		&\left\|\bm c_l[n]-\bm c_l[n-1]\right\|_2\leqslant V_{max}\Delta_t, \forall l,n,\label{c3}\\
		&R_l[n]\geqslant R_{min},\forall l,n, \text{(\ref{SenCon})}, \label{c4}
	\end{align}\label{op0}%
\end{subequations}
in which (\ref{c1}) is the transmit power constraint and $P_t$ is the maximum power of the ISAC base station, (\ref{c2}) imposes the constant modulus constraint for the analog beamformer $F_{RF}[n]$. (\ref{c3}) ensures the velocity of each legitimate UAV does not exceed the maximum velocity $V_{max}$, and (\ref{c4}) is the quality of service (QoS) requirement for legitimate UAVs and the sensing performance constraint.

It can be concluded that (\ref{op0}) is a high-dimensional, non-convex problem due to several intrinsic challenges. Firstly, the constant-modulus constraint (\ref{c2}) and the limited radio-frequency chains $N_{\text{RF}} \ll N_t$ restrict spatial degrees of freedom, while the functional separation of $\bm F_{BB}[n]$ and $\bm w[n]$ further fragments the available resources. Secondly, the optimization variables are tightly coupled with each other. For example, designing hybrid beamformer gains depends on channel gains, which is contingent upon legitimate UAV positions. Besides, the transmit power constraint, QoS constraint, and sensing performance constraint are all non-convex. Finally, the time-varying channels and trajectory dynamics necessitate sequential decision-making with foresight. These properties collectively render the problem NP-hard and unsuitable for conventional convex optimization or block coordinate descent methods, which often converge to poor local optima and fail to capture temporal correlations.

Even when relaxing the hardware constraints by considering fully-digital beamforming as a precursor to HBF, the optimization difficulty persists significantly. The full-digital beamforming problem remains high-dimensional and non-convex due to several intrinsic complexities: the need for joint optimization of $\bm F_{dig}=\bm F_{RF}\bm F_{BB}$ and $\bm w_{dig}=\bm F_{RF}\bm w$ under stringent constraints, the coupling between beamformer designs and legitimate UAV positions, and the non-convex nature of optimization constraints. These persistent challenges motivate the need for advanced optimization techniques capable of handling such complex design spaces. DRL offers unique advantages for solving such non-convex sequential optimization problems, which align precisely with the core requirements of optimization problem (\ref{op0}).

We therefore propose a two-stage optimization framework that strategically leverages DRL while circumventing its limitations in handling intricate physical constraints. Detailed design is elaborated in Section IV.

\section{Two-Stage Optimization Framework}
In the first stage, a DRL agent is employed to obtained the fully-digital beamformers $\bm F_{dig}$ and $\bm w_{dig}$. This approach effectively bypasses the mathematical intractability of the original problem while maintaining performance excellence. The second stage then decomposes the high-performance full-digital solution into practical hybrid components $\bm F_{RF}$, $\bm F_{BB}$, and $\bm w$ via matrix factorization techniques with low complexity, ensuring adherence to the constant-modulus and transmit power constraints inherent in hybrid architectures.
\subsection{PPO-Based Joint Digital Beamforming and Trajectory Design}
The agent first interact with the environment to observe the state $s_n$, which is input to the actor network $\pi_\vartheta(a_n|s_n)$ to decide the action $a_n$ to execute. The critic network $V_\varphi(s_n)$ evaluates state values to guide policy updates through clipped objective optimization, ensuring stable and efficient learning. For the digital beamforming and trajectory optimization part, the sum secrecy maximization problem (\ref{op0}) is firstly formulated as a Markov decision process (MDP), and the core components are proposed as follows.
\subsubsection{State Space $\mathcal{S}$}
As the ISAC base station requires to determine the transmit beamforming according to real-time channel gain, and trajectories of the legitimate UAVs depend on its previous position and the locations of other nodes, the state at step $n$ is defined as\vspace{-0.3em}
\begin{equation}
	s_n=\{\bm H_L[n],\bm H_E[n], \bm C_L[n-1],\bm c_A, \bm C_E\},\vspace{-0.3em}
\end{equation}
where $\bm H_L=\{\bm h_1,\ldots,\bm h_L\}$ and $\bm H_E=\{\bm h_1,\ldots,\bm h_E\}$ is the combined legitimate channel and eavesdropping channel gains, $\bm C_L$ is the locations of legitimate UAVs, $\bm c_A$ and $\bm C_E$ are the locations of the base station and eavesdropper UAVs respectively. It is noteworthy that prior to being fed into neural networks, $\bm H_L$ and $\bm H_E$ must undergo real-imaginary decomposition.
\subsubsection{Action Space $\mathcal{A}$}
Recall the variables of (\ref{op0}), the action at step $n$ must contain the digital beamformer and the movements of legitimate UAVs, that is\vspace{-0.3em}
\begin{equation}
	a_n=\{\bm F[n],\bm w[n],\bm\delta_x[n],\bm\delta_y[n]\},\vspace{-0.3em}
\end{equation}
where $\bm\delta_x$ and $\bm\delta_y$ are the movement of legitimate UAVs along the $x$-axis and $y$-axis. Actually, the actor network only output real values for $\bm F$ and $\bm w$, which are subsequently combined and normalized to complex matrices.

\subsubsection{Reward $\mathcal{R}$}
The total reward consists three parts as $R_n=R_{com}[n]+R_{sen}[n]+R_{QoS}[n]$. $R_{com}[n]$ is the secure communication part, which equals the sum secrecy rate of all legitimate UAVs at step $n$. $R_{sen}[n]$ and $R_{QoS}[n]$ are the sensing performance penalty and QoS penalty respectively\vspace{-0.3em}
\begin{subequations}
	\begin{equation}
		R_{sen}[n]=-\textstyle\sum_{e=1}^E\left[\Gamma_e-P_{\phi_,\theta_e}[n]d_e^{-2}\right]^+,\vspace{-0.3em}
	\end{equation}
	\begin{equation}
		R_{QoS}[n]=-\textstyle\sum_{l=1}^L\left[R_{min}-R_{s,l}[n]\right]^+.\vspace{-0.3em}
	\end{equation}
\end{subequations}

As such, the DRL agent is encouraged to optimize the sum secrecy rate, while concomitantly ensuring the sensing performance and QoS constraints.

For each episode of PPO training, the replay buffer is first reset and then repopulated with data generated by the current policy. Based on the data extract from the replay buffer, the objective function for actor and critic network updates are\vspace{-0.3em}
\begin{subequations}
	\begin{equation}
		J(\vartheta)\!=\!\mathbb E\left[\min\left(r(\vartheta)A_n,{\rm clip}\left(r(\vartheta),1\!-\!\epsilon,1\!+\!\epsilon\right) A_n\right)\right],\vspace{-0.3em}
	\end{equation}
	\begin{equation}
		L(\varphi)\!=\!\mathbb E[(V_\varphi(s_n)-\hat R_n)^2],\vspace{-0.3em}
	\end{equation}
\end{subequations}
where $r(\vartheta)=\frac{\pi_\vartheta(a_n|s_n)}{\pi_{\vartheta_{old}}(a_n|s_n)}$ is the probability ratio, $A_n$ is the advantage function estimated via generalized advantage estimation (GAE), $\epsilon$ is the clip hyperparameter, $\hat R_n$ is the discounted return.

\subsection{Power-Constrained Hybrid Decomposition via AO}
After the fully-digital beamformer $\bm F_{opt}$ and $\bm w_{opt}$ are constructed by PPO, an AO method is employed to decompose them into the analog beamformer $\bm F_{RF}$ and digital beamformers $\bm F_{BB}$ and $\bm w$. The objective of decomposition is to minimize the difference between the fully-digital and hybrid beamformers\vspace{-0.3em}
\begin{equation}
	\underset{\bm F_{RF},\bm F_{BB}, \bm w}{\mathop{\text {min}}}\left\|\bm F_{opt}-\bm F_{RF}\bm F_{BB}\right\|_F^2+\left\|\bm w_{opt}-\bm F_{RF}\bm w\right\|_2^2,\label{op2}\vspace{-0.3em}
\end{equation}
with the power constraint (\ref{c1}) and the constant modulus constraint (\ref{c2}) guaranteed. For solving (\ref{op2}), an AO based algorithm is developed, and the decomposition process consists of the following three key steps:
\subsubsection{Digital beamformers Update}
After the initial beamformers $\bm F_{RF}^{(0)}$, $\bm F_{BB}^{(0)}$, and $\bm w^{(0)}$ are acquired, we firstly optimize $\bm F_{BB}$ and $\bm w$ while the analog beamformer $\bm F_{RF}$ is fixed. This reduces to a conventional least-squares estimation problem that minimizes the approximation error between the optimal full-digital solution and its hybrid counterpart, leading to a closed-form solution through the Moore-Penrose pseudo-inverse\vspace{-0.3em}
\begin{subequations}\label{digital}
	\begin{equation}
		\bm F_{BB}^{(1)}=\bm F_{RF}^{(0)\dagger}\bm F_{opt},\vspace{-0.3em}
	\end{equation}
	\begin{equation}
		\bm w^{(1)}=\bm F_{RF}^{(0)\dagger}\bm w_{opt}.\vspace{-0.3em}
	\end{equation}
\end{subequations}

This solution represents the minimum-norm least-squares approximation that optimally projects the full-digital beamformers onto the subspace spanned by the analog beamformer.

\subsubsection{Analog beamformer Update}
The optimization of the analog beamformer $\bm F_{RF}$ is fundamentally governed by the constant modulus constraint (\ref{c2}), which arises from the use of phase shifters in HBF architectures. The core principle underlying the analog beamformer update is correlation maximization between the optimal fully-digital beamformer and the hybrid approximation. For each element of $\bm F_{RF}^{(1)}$, the optimal phase $\psi_{i,j}$ is derived by maximizing the correlation between the fully-digital and HBF structures. Specifically, the update rule is\vspace{-0.3em}
\begin{equation}
	\psi_{i,j}=\angle\left(\bm F_{opt}[i,:]\bm F_{BB}^{(1)H}[j,:]+\bm w_{opt}[i,:]\bm w^{(1)H}[j,:]\right),\label{analog}\vspace{-0.3em}
\end{equation}
then we could update $\bm F_{RF}^{(1)}(i,j)$ as $\bm F_{RF}^{(1)}(i,j)=1/\sqrt{N_t}e^{j\psi_{i,j}}$.

\subsubsection{Power Normalization}
The power normalization step ensures strict adherence to the total transmit power constraint (\ref{c1}), which involves calculating the total radiated power and applying appropriate scaling. For optimized hybrid beamformers, the total power consumption is computed as
\begin{equation}
	P_{total}=||\bm F_{RF}^{(k)}\bm F_{BB}^{(k)}||_F^2+\|\bm F_{RF}^{(k)}\bm w^{(k)}\|_2^2,
\end{equation}
and the scaling factor is calculated as $\eta=\sqrt{{P_{t}}/{P_{total}}}$, then the digital beamformers are then scaled according to $\bm F_{BB}^{(k)}=\eta\bm F_{BB}^{(k)}$ and $\bm w^{(k)}=\eta\bm w^{(k)}$.

The proposed HBF decomposition is achieved through an iterative AO framework that cyclically executes these three fundamental steps, which progressively minimizes the approximation error between the optimal full-digital solution and its hybrid counterpart while maintaining all constraints.
\section{Simulation Experiments}
The ISAC base station is assumed to be located at the origin, and the height of its antenna array is neglected. The considered multi-UAV ISAC network comprises $L=\text{4}$ legitimate UAVs, with initial locations at $[\text{20},\text{80},\text{20}]^T$, $[\text{20},\text{70},\text{15}]^T$, $[\text{70},\text{30},\text{30}]^T$, and $[\text{80},\text{10},\text{15}]^T$ respectively. Meanwhile, $E=\text{3}$ eavesdropper UAVs are positioned near the legitimate UAVs at locations $[\text{20},\text{40},\text{20}]^T$, $[\text{50},\text{50},\text{30}]^T$, and $[\text{80},\text{70},\text{40}]^T$. The UPA consists $N_t=8\times8$ antennas, and the number of RF chains is $N_{RF}=8$. The transmit power of ISAC base station is $P_t=10\si{\watt}$ and the loss component of channel gain is $\kappa=1.8$. With time interval $\Delta_t=0.5\si{\second}$ and total transmission duration $T=100\si{\second}$, we have the number of time slots $N=200$. Sensing performance threshold is set as $\Gamma_e=0.5e^{-5}$ and the QoS requirement $R_{min}$ is 0.5 bps/Hz. For all UAVs the noise power is $\sigma_i^2=1e^{-13}$.

For the training hyperparameters for PPO, the batch size is set to $1e3$ and the mini-batch size is $200$, the learning rate of actor and critic network is $1e^{-4}$ and $3e^{-4}$ respectively. The discount factor is $\gamma=0.9$ and the clip parameter is $\epsilon=0.2$.

\begin{figure}[h]
	\centering
	\includegraphics[width=2in]{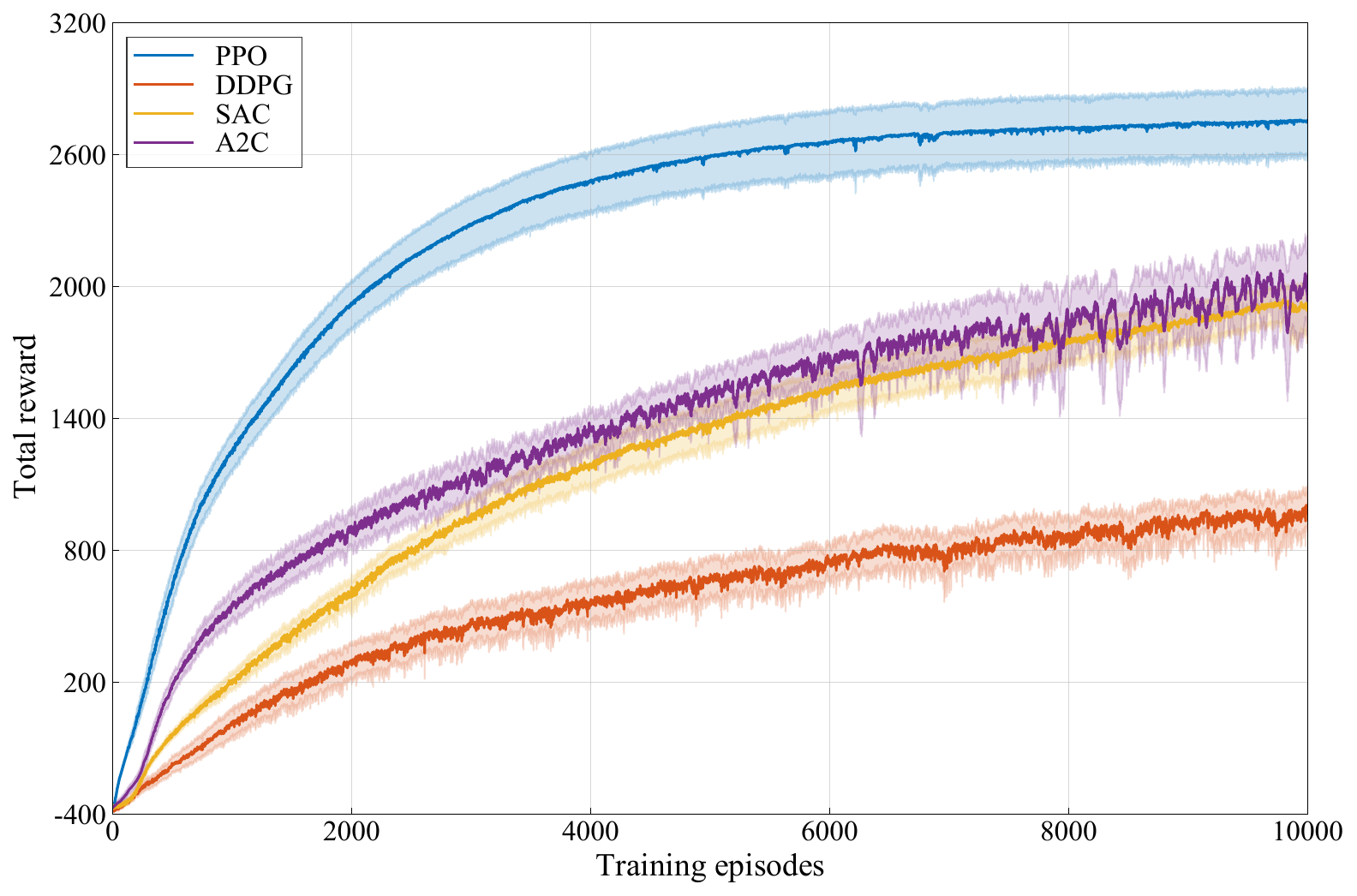}
	\caption{Convergence performance comparison of PPO-based algorithm with DDPG, SAC, and A2C.}
	\vspace{-0.3em}
	\label{fig2}
\end{figure}

The convergence performance of the proposed PPO-based algorithm for optimizing fully-digital beamformer and trajectories of legitimate UAVs is illustrated in Fig. \ref{fig2}. Three algorithms are taken as benchmarks, \ie, Deep Deterministic Policy Gradient(DDPG), Soft Actor-Critic (SAC), and Advantage Actor-Critic (A2C). The maximum episode of all algorithms is set to $1e^4$ and each curve in Fig. \ref{fig2} is compounded by 20 tracks corresponding to 20 random seeds. Fig. \ref{fig2} reveals that the PPO-based algorithm demonstrates superior total return and enhanced training stability compared to alternative methods, indicating its effectiveness in jointly optimizing digital beamforming and UAV trajectories to achieve a higher sum secrecy rate.

\begin{figure}[t]
	\centering
	\includegraphics[width=2.6in]{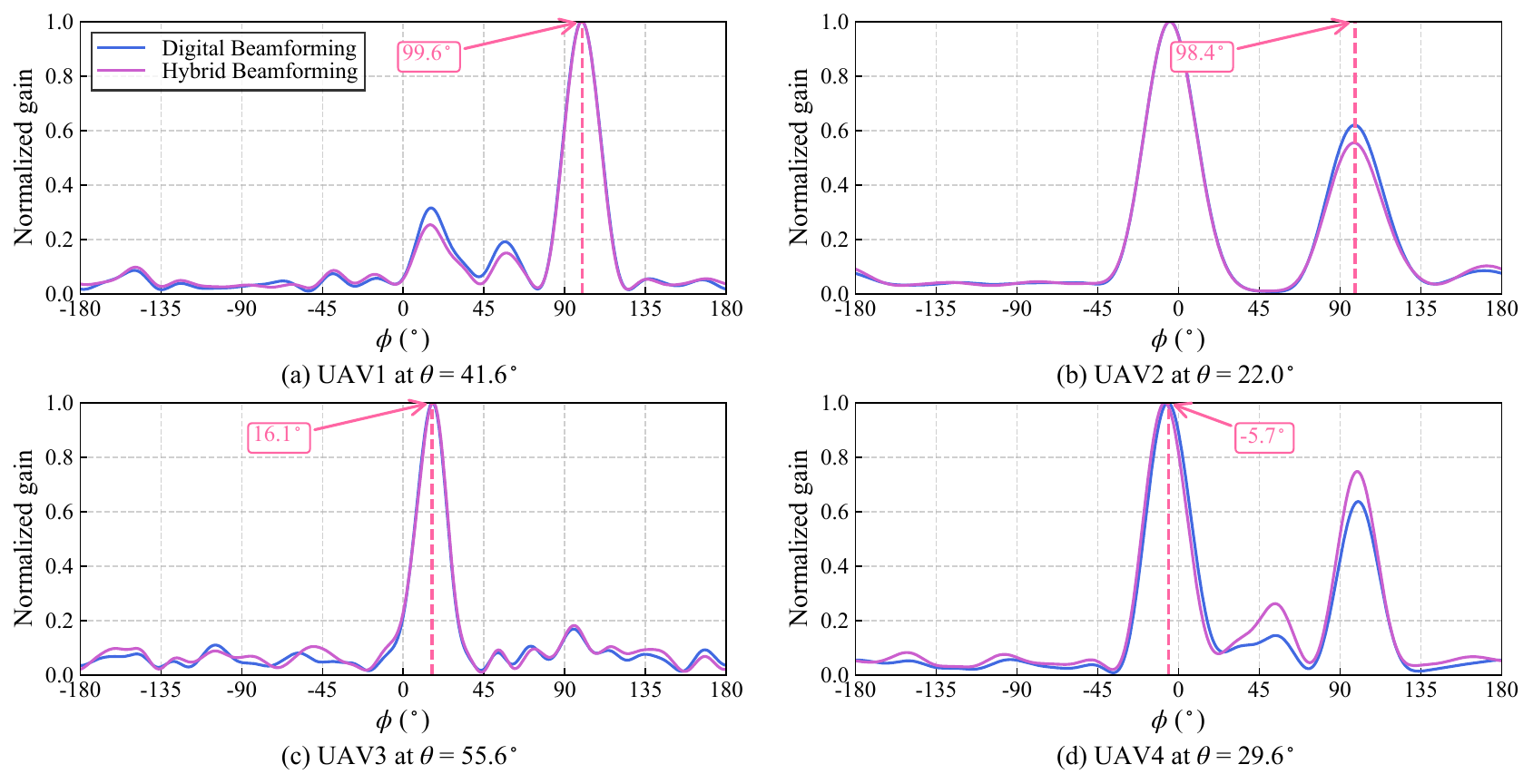}
	\caption{Beam pattern comparison of fully-digital and HBF.}
	\vspace{-1.5em}
	\label{fig3}
\end{figure}

For validating the effectiveness of the optimized beamforming and accessing performance gap between fully-digital beamforming and HBF, the transmit beampatterns of both schemes in time slot 40 are compared in Fig. \ref{fig3}. Each subfigure displays the azimuth cut corresponding to each legitimate UAV, with the respective elevation cut marked. We can conclude from Fig. \ref{fig3} that the proposed algorithm demonstrates flexible and precise steering of the main lobe toward each legitimate UAV's direction. Moreover, the high degree of similarity between the digital and HBF beampatterns confirms the effect of the AO-based hybrid decomposition approach.

\begin{figure}[ht]
	\centering
	\includegraphics[width=2.10in]{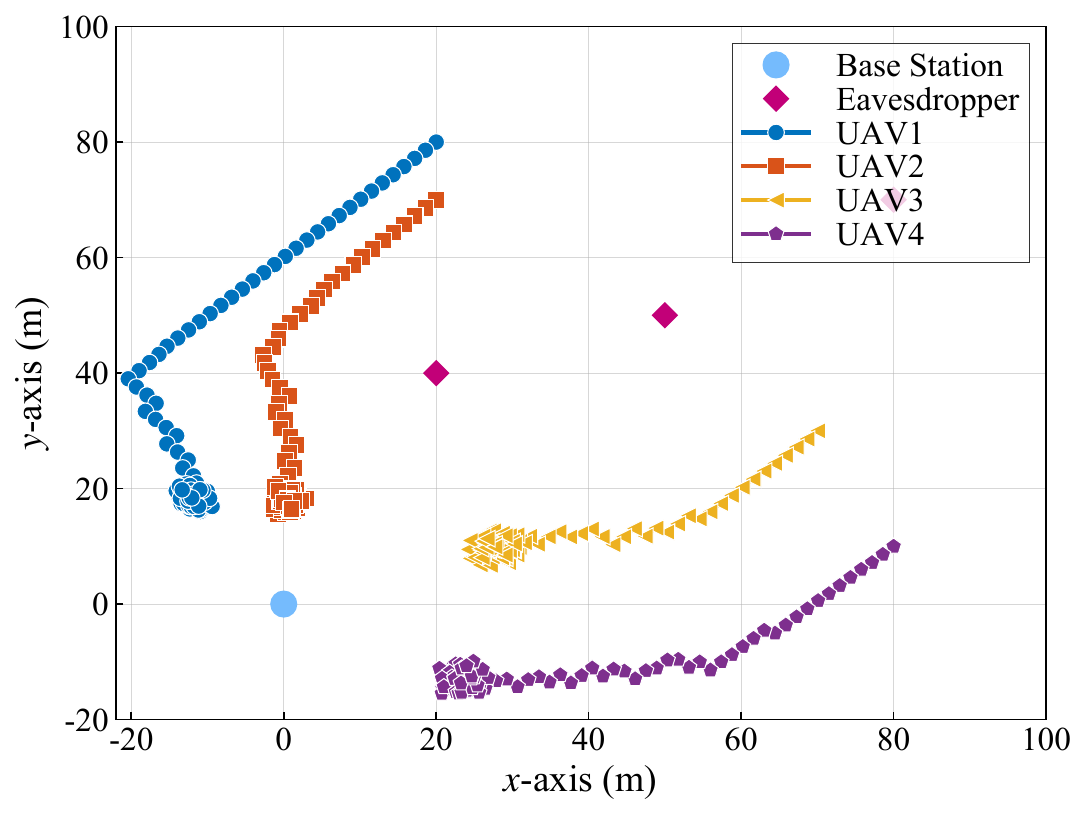}
	\caption{Optimized trajectory of legitimate UAVs.}
	\vspace{-1.0em}
	\label{fig4}
\end{figure}

Fig. \ref{fig4} depicts the optimized trajectories of legitimate UAVs, showing the base station (blue circle), eavesdropper UAVs (pink diamond). Each legitimate UAV dynamically navigates toward an optimal hovering point near the base station to maximize secrecy rate. Throughout their flight, the UAVs consistently maneuver to maximize their distance from the eavesdropper UAVs, thereby enhancing security performance.
\vspace{-1em}

\begin{figure}[ht]
	\centering
	\includegraphics[width=2.00in]{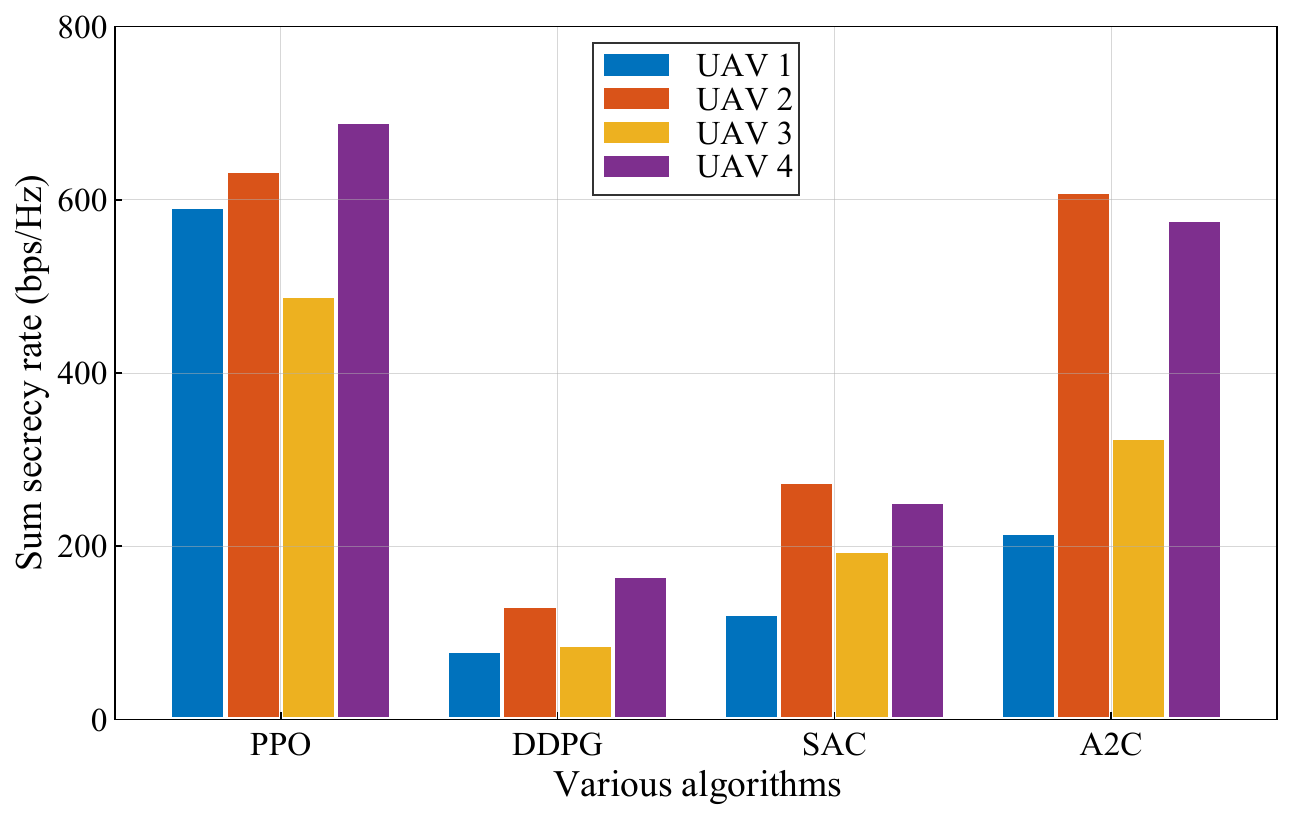}
	\caption{Comparison of sum secrecy rate of each legitimate UAV.}
	\vspace{-1.0em}
	\label{fig5}
\end{figure}

Fig. \ref{fig5} compares the sum secrecy rate achieved by legitimate UAVs with HBF. As evidenced by Fig. \ref{fig5}, the proposed PPO-based algorithm demonstrates significantly higher secrecy rates across all UAVs compared with other algorithms. The relatively balanced distribution of secrecy rates among UAVs under PPO also reflects better fairness performance. Specifically, compared to the best-performing alternative approach, which is A2C, the PPO-based algorithm achieves a 39.3\% improvement in total secrecy rate.
\vspace{-0.8em}


\section{Conclusion}
The secure communication of a multi-UAV communication network with ISAC was conducted in this paper, where a dual-function terrestrial base station transmits confidential information to multiple legitimate UAVs and AN to interference multiple eavesdropper UAVs. The installation of a large scale antenna array at the ISAC base station was undertaken with the objective of enhancing transmission efficiency and improving spatial resolution of the beam. To this end, a HBF architecture was employed. In the purpose of maximizing the sum secrecy rate over all legitimate UAVs, the hybrid beamformer in conjunction with the trajectory of legitimate UAVs were jointly optimized, and a two-stage optimization framework was developed. Specifically, the DRL agent efficiently handles the computationally intensive full-digital optimization, while deterministic decomposition algorithms ensure physical realizability and precise performance-complexity trade-offs. Simulation results demonstrate that the proposed algorithm achieves a superior performance compared to benchmark schemes, enhancing the sum secrecy rate by 39.3\%.

\bibliographystyle{IEEEtran}
\bibliography{myref}

@article{RN1142,
   author = {Li, S. and Dong, H. and Shan, C. and Fang, X. and Wu, W. and Li, Z.},
   title = {Secure Hybrid Beamforming Design for mm{W}ave Integrated Sensing and Communication Systems},
   journal = {IEEE Transactions on Vehicular Technology},
   volume = {74},
   number = {7},
   pages = {10622-10638},
   ISSN = {1939-9359},
   DOI = {10.1109/TVT.2025.3544397},
   year = {2025},
   month = {Jul.},
   type = {Journal Article}
}

@article{RN948,
   author = {Dong, R. and Wang, B. and Cao, K. and Tian, J. and Cheng, T.},
   title = {Secure Transmission Design of RIS Enabled {UAV} Communication Networks Exploiting Deep Reinforcement Learning},
   journal = {IEEE Transactions on Vehicular Technology},
   volume = {73},
   number = {6},
   pages = {8404-8419},
   ISSN = {1939-9359},
   DOI = {10.1109/TVT.2024.3357821},
   year = {2024},
   month = {Jun.},
   type = {Journal Article}
}

@article{RN1144,
   author = {Sobhi-Givi, S. and Nouri, M. and Shayesteh, M. G. and Kalbkhani, H. and Ding, Z.},
   title = {Joint Power Allocation and User Fairness Optimization for Reinforcement Learning Over mm{W}ave-{NOMA} Heterogeneous Networks},
   journal = {IEEE Transactions on Vehicular Technology},
   volume = {73},
   number = {9},
   pages = {12962-12977},
   ISSN = {1939-9359},
   DOI = {10.1109/TVT.2024.3386587},
   year = {2024},
   month = {Sept.},
   type = {Journal Article}
}

@article{RN1104,
   author = {Zhang, Z. and Wang, J. and Chen, J. and Fu, H. and Tong, Z. and Jiang, C.},
   title = {Diffusion-Based Reinforcement Learning for Cooperative Offloading and Resource Allocation in Multi-{UAV} Assisted Edge-Enabled Metaverse},
   journal = {IEEE Transactions on Vehicular Technology},
   volume = {74},
   number = {7},
   pages = {11281-11293},
   ISSN = {1939-9359},
   DOI = {10.1109/TVT.2025.3544879},
   year = {2025},
   month = {Jul.},
   type = {Journal Article}
}

@article{RN1036,
   author = {Ju, Y. and Gao, Z. and Wang, H. and Liu, L. and Pei, Q. and Dong, M. and Mumtaz, S. and Leung, V. C. M.},
   title = {Energy-Efficient Cooperative Secure Communications in mm{W}ave Vehicular Networks Using Deep Recurrent Reinforcement Learning},
   journal = {IEEE Transactions on Intelligent Transportation Systems},
   volume = {25},
   number = {10},
   pages = {14460-14475},
   ISSN = {1558-0016},
   DOI = {10.1109/TITS.2024.3394130},
   year = {2024},
   month = {Oct.},
   type = {Journal Article}
}

@article{RN1139,
   author = {Geng, Y. and Cheng, T. Hiang and Zhong, K. and Teh, K. Chan and Wu, Q.},
   title = {Joint Beamforming for {CRB}-Constrained {IRS}-Aided {ISAC} System via Product Manifold Methods},
   journal = {IEEE Transactions on Wireless Communications},
   volume = {24},
   number = {1},
   pages = {691-705},
   ISSN = {1558-2248},
   DOI = {10.1109/TWC.2024.3498105},
   year = {2025},
   month = {Jan.},
   type = {Journal Article}
}

@article{RN1095,
   author = {Ye, J. and Dai, J. and Pan, C. and Wang, K. and Li, J.},
   title = {Joint Active and Passive Beamforming Design for Secure {RIS}-Aided {ISAC} System},
   journal = {IEEE Wireless Communications Letters},
   volume = {14},
   number = {3},
   pages = {916-920},
   ISSN = {2162-2345},
   DOI = {10.1109/LWC.2025.3528080},
   year = {2025},
   month = {Mar.},
   type = {Journal Article}
}

@article{RN1094,
   author = {Li, H. and Xiao, M. and Wang, K. and Kim, D. I. and Debbah, M.},
   title = {Large Language Model Based Multi-Objective Optimization for Integrated Sensing and Communications in {UAV} Networks},
   journal = {IEEE Wireless Communications Letters},
   volume = {14},
   number = {4},
   pages = {979-983},
   ISSN = {2162-2345},
   DOI = {10.1109/LWC.2025.3529082},
   year = {2025},
   month = {Apr.},
   type = {Journal Article}
}

@article{RN1087,
   author = {Tang, Y. and Zhu, G. and Xu, W. and Cheung, M. H. and Lok, T. M. and Cui, S.},
   title = {Integrated Sensing, Computation, and Communication for {UAV}-assisted Federated Edge Learning},
   journal = {IEEE Transactions on Wireless Communications},
   volume = {24},
   number = {4},
   pages = {2647-2662},
   ISSN = {1558-2248},
   DOI = {10.1109/TWC.2024.3523381},
   year = {2025},
   month = {Apr.},
   type = {Journal Article}
}

@article{RN1028,
   author = {Jing, X. and Liu, F. and Masouros, C. and Zeng, Y.},
   title = {ISAC from the Sky: {UAV} Trajectory Design for Joint Communication and Target Localization},
   journal = {IEEE Transactions on Wireless Communications},
   pages = {12857-12872},
   ISSN = {1558-2248},
   DOI = {10.1109/TWC.2024.3396571},
   year = {2024},
   month = {Oct.},
   type = {Journal Article}
}

@article{RN873,
   author = {Lyu, Z. and Zhu, G. and Xu, J.},
   title = {Joint Maneuver and Beamforming Design for {UAV}-Enabled Integrated Sensing and Communication},
   journal = {IEEE Transactions on Wireless Communications},
   volume = {22},
   number = {4},
   pages = {2424-2440},
   ISSN = {1558-2248},
   DOI = {10.1109/TWC.2022.3211533},
   year = {2023},
   month = {Apr.},
   type = {Journal Article}
}

@article{RN1132,
   author = {Yao, J. and Yang, Z. and Yang, Z. and Xu, J. and Quek, T. Q. S.},
   title = {{UAV}-Enabled Secure {ISAC} Against Dual Eavesdropping Threats: Joint Beamforming and Trajectory Design},
   journal = {IEEE Wireless Communications Letters},
   volume = {14},
   number = {10},
   pages = {3199-3203},
   ISSN = {2162-2345},
   DOI = {10.1109/LWC.2025.3588758},
   year = {2025},
   month = {Oct.},
   type = {Journal Article}
}

@article{RN1029,
   author = {Zhang, J. and Xu, J. and Lu, W. and Zhao, N. and Wang, X. and Niyato, D.},
   title = {Secure Transmission for {IRS}-Aided {UAV}-{ISAC} Networks},
   journal = {IEEE Transactions on Wireless Communications},
   volume = {23},
   number = {9},
   pages = {12256-12269},
   ISSN = {1558-2248},
   DOI = {10.1109/TWC.2024.3390169},
   year = {2024},
   month = {Sept.},
   type = {Journal Article}
}
\end{document}